\documentclass[RNAAS]{aastex62}

\usepackage{animate}

\graphicspath{{./}{figures/}}

\begin{document}

\title{Photometric Observations of the 2017 Outburst of Recurrent Nova M31N 2007-10b}

\author[0000-0002-7434-0863]{Quentin J. Socia}
\affiliation{Department of Astronomy, San Diego State University, San Diego, CA 92182, USA}

\author{Martin Henze}
\affiliation{Department of Astronomy, San Diego State University, San Diego, CA 92182, USA}

\author{Allen W. Shafter}
\affiliation{Department of Astronomy, San Diego State University, San Diego, CA 92182, USA}

\author{J. Chuck Horst}
\affiliation{Department of Astronomy, San Diego State University, San Diego, CA 92182, USA}

\keywords{Galaxies: individual: M31 ---  novae, cataclysmic variables  --- stars: individual: M31N 2007-10b, M31N 2017-12a}

\section{} 

M31 is an ideal laboratory for observing and studying recurrent novae. To date, there have been 18 recurrent nova discovered in M31, six of which have recurrence periods less than nine years (\citealt{2015ApJS..216...34S}; \citealt{2018ApJ...857...68H}).

M31N 2017-12a (AT2017jdm) is a transient that was reported near the center of M31 \citep{2017TNSTR1467....1B}, with the time of outburst estimated to be 2017 December $24.141\pm{0.42}$ based on constraints from previous data \citep{2017ATel11091....1T}. Here we present the photometric observations of M31N 2017-12a, which we confirm as a recurrence of M31N 2007-10b \citep{2017ATel11088....1W}. 

Observations of M31N 2017-12a were conducted using the Mount Laguna Observatory 1.0m telescope. The Johnson--Cousins filters $B, V,$ and $R$ were used for all four nights of observation. Standard bias images and twilight flat-fields were acquired for each night. On 2017 December 25 and 2017 December 26, a 360 second exposure time was utilized for all three filters with $\sim$ 30 seconds in between each image for readout, filter cycling, and computer overhead. On 2017 December 27 and 2017 December 28, the exposure times were increased to 600 seconds. Cloudy weather limited the number of images taken on the third night.

Each image was reduced with standard over-scan subtraction, bias subtraction, and flat-field normalization using custom python subroutines in conjunction with astropy \citep{2018arXiv180102634T}. WCS solutions for each frame were obtained by using the astrometry.net solve-field package, which uses the USNO-B and 2MASS catalogs for its index files \citep{2010AJ....139.1782L}. Aperture photometry and JD to BJD correction was performed using AstroImageJ (AIJ) \citep{2017AJ....153...77C}. The differential photometry utilized various stars in the 13.5 square arc-minute field of view. 

\begin{figure}[ht]
\centering
\begin{center}
\gridline{\fig{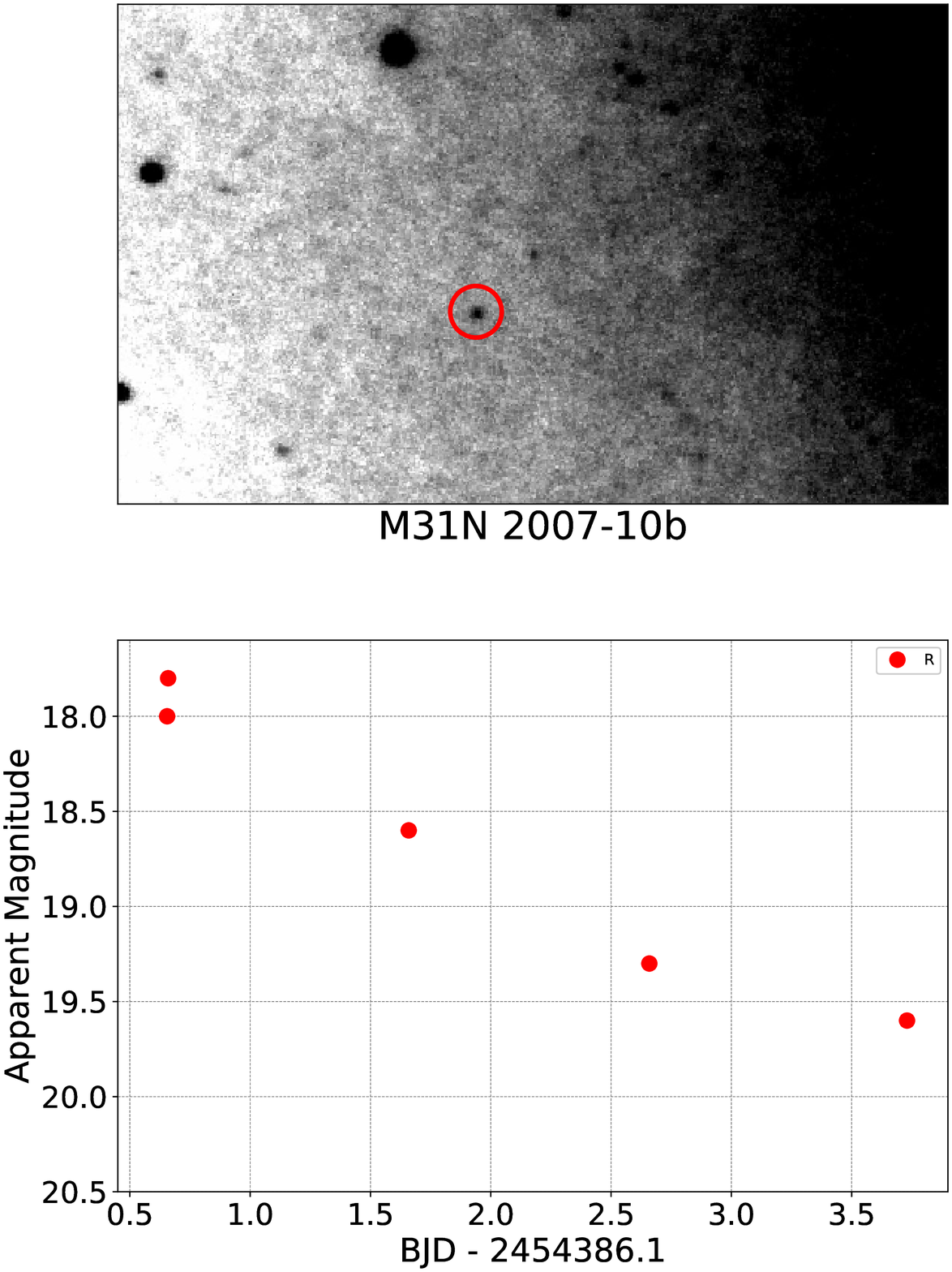}{0.33\textwidth}{(a)}
          \fig{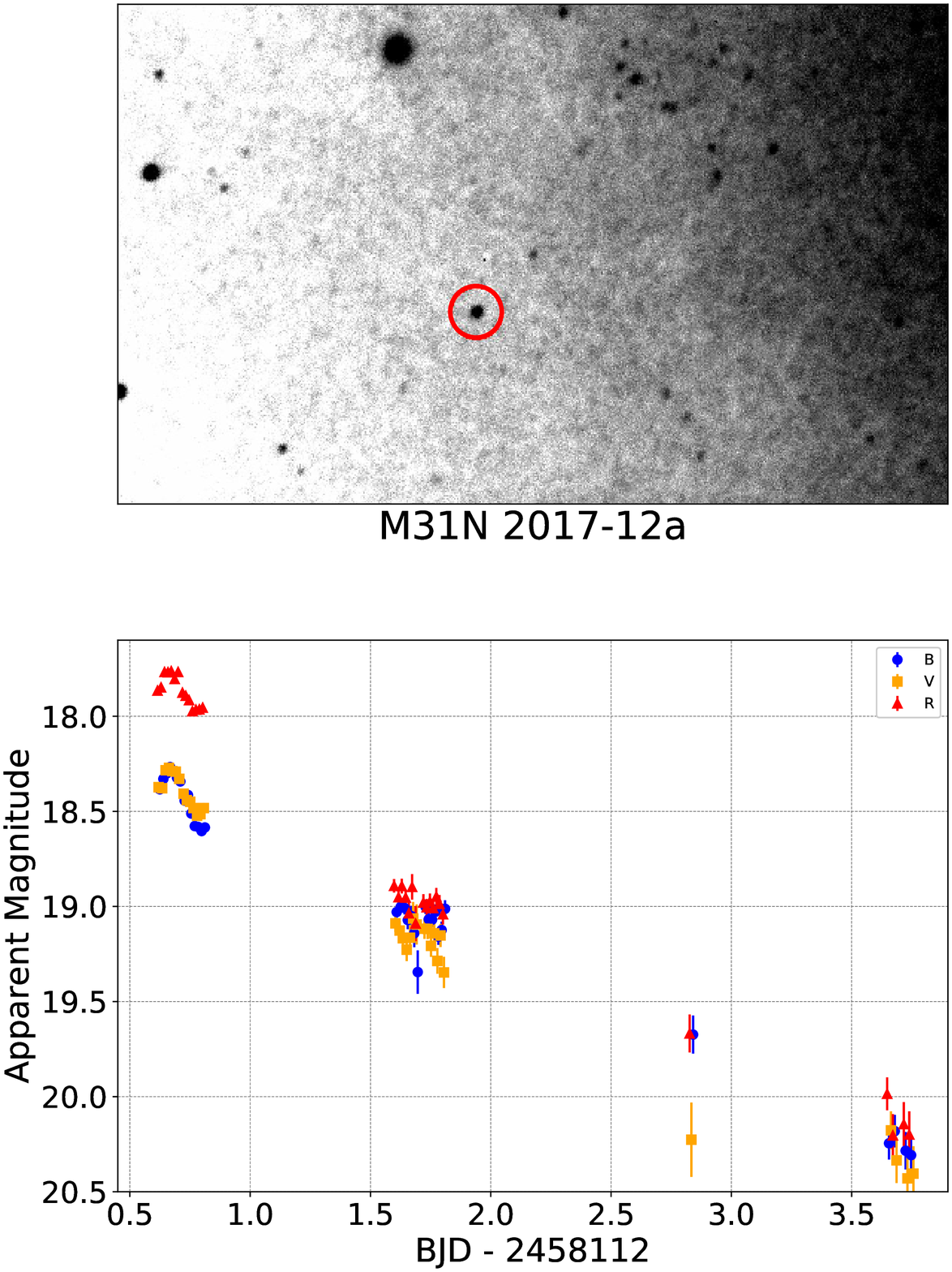}{0.33\textwidth}{(b)}
          \fig{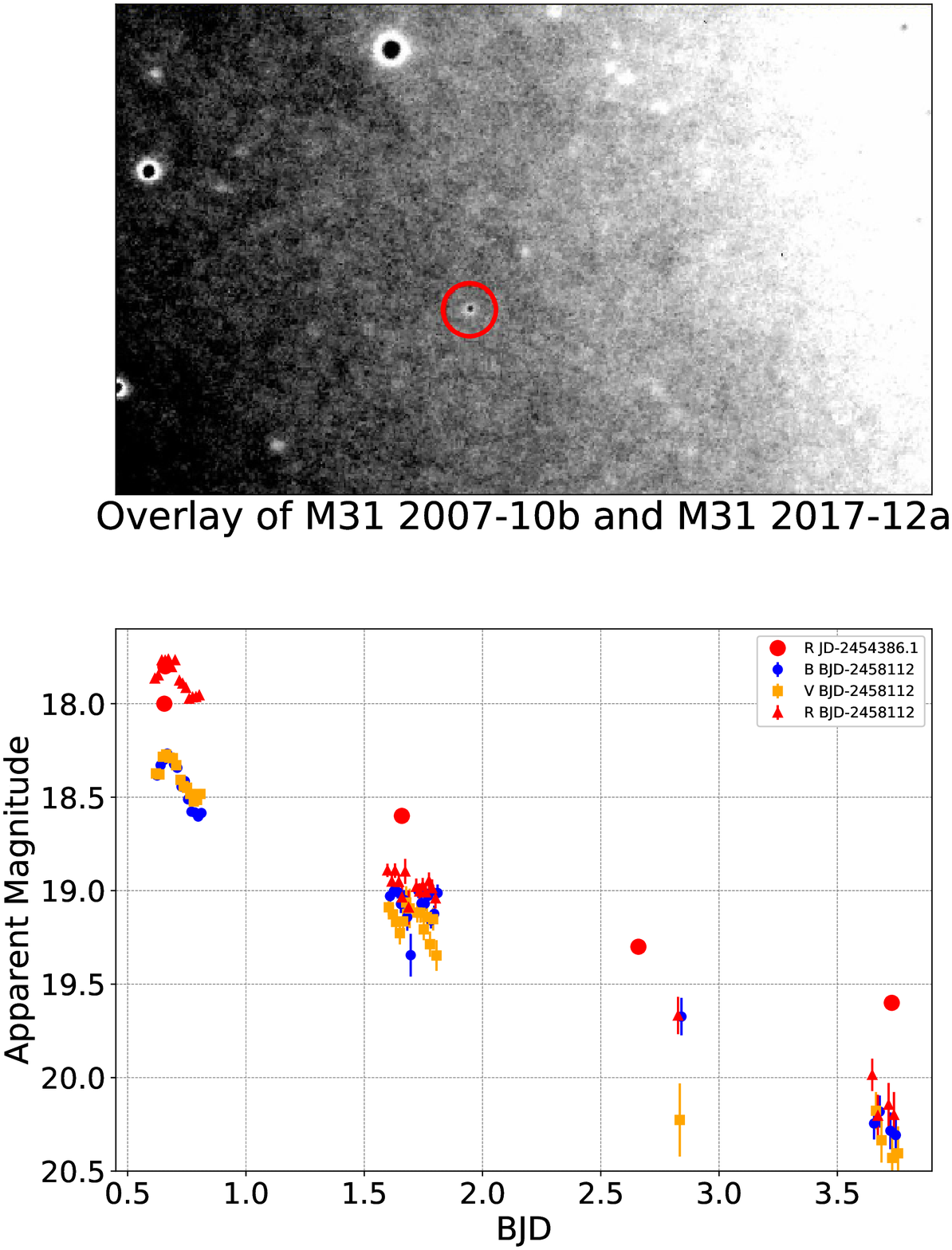}{0.33\textwidth}{(c)}}
\caption{\textit{Top:} The eruption of M31N 2007-10b taken in October of 2007 by SuperLOTIS, just outside the M31 Bulge, followed by the eruption of M31N 2017-12a, taken by the Liverpool Telescope on 2017 December 24 in the same position shortly after the 2017 outburst detection, and the last image is an overlay with both images with the SuperLOTIS inverted in color. \textit{Bottom:} $BVR$ photometry from MLO 1m telescope from 2018 December 25 to 2018 December 28 and R photometry from SuperLOTIS in 2007 October. The large circles are SuperLOTIS data points.
}
\label{fig:1}
\end{center}
\end{figure}

We measure the position of the nova to be $\alpha_{2000} = 00^{h}43^{m}29^{s}.42, \delta_{2000} = 43^{o}17'13 \:  45''$ with an average uncertainty of about 1 arcsec. We find the peak magnitudes to be $B = 18.27 \pm 0.02$; $V = 18.27\pm 0.03$, $R = 17.76 \pm 0.02$, with the peaks occurring at BJD = 2458112.6674, 2458112.6629, 2458112.6719 respectively. Using a linear fit to the data, we find the time to fade 2 magnitudes is $2.61\pm{0.05}$ days in the $R$ filter, $2.81\pm{0.05}$ days in the $V$ filter, and $3.36\pm{0.04}$ days in the $B$ filter. The bottom panels of Figure \ref{fig:1} show the $BVR$ light curves of all four observing nights and light curves from SuperLOTIS in 2007 (\citealt{2007ATel.1238....1B}; \citealt{2007ATel.1242....1R}).

The discovery location of M31N 2007-10b ($\alpha_{2000} = 00^{h}43^{m}29^{s}.48, \delta_{2000} = 43^{o}17'13 \; 50''$; \citealt{2007ATel.1238....1B}) places it 0.9 arc-seconds from the position of the 2017 nova, within the uncertainty of our measurement. The 2007 discovery was made in the $R$ filter at a magnitude of $R = 18.0$ and $R = 17.8$. Based on observational constraints, it is estimated to have occurred at 2007 October $12.827\pm{0.44}$, and faded 2 magnitudes in roughly 3.5 days \citep{2007ATel.1242....1R}. The left panels of Figure \ref{fig:1} show comparison images of both novae, one taken by SuperLOTIS \citep{2008AIPC.1000..535W} in October 2007 and the other by the Liverpool Telescope on 2017 December 24. The top panels of Figure \ref{fig:1} show images from both epochs carefully overlaid with each other.

Given that the two eruptions are spatially coincident within the uncertainties of the measurements, coupled with their similarity in peak brightness and fade rate, we conclude that they were caused by the same nova progenitor system. This discovery adds another crucial member to the rare phenomena of short period recurrent novae, and another candidate for the possible Type Ia supernova precursor population.

\end{document}